\documentclass[twocolumn,showpacs,preprintnumbers,amsmath,amssymb]{revtex4}

\usepackage{epsfig}
\usepackage{dcolumn}
\usepackage{bm}

\begin{document}

\preprint{}

\title{Nonequilibrium many-body dynamics along a dissipative Hubbard
chain:\break Symmetries and Quantum Monte Carlo
simulations}

\author{L.~M\"uhlbacher$^1$}
\affiliation{${}^1$School of Chemistry, The Sackler Faculty of Exact Science,
 Tel Aviv University, Tel Aviv 69978, Israel}
\author{J.~Ankerhold$^2$}
\affiliation{${}^2$Physikalisches Institut,
 Albert-Ludwigs-Universit\"at, D-79104 Freiburg, Germany }

\date{\today}

\begin{abstract}
  The nonequilibrium dynamics of correlated charge transfer along a
  one-dimensional chain in presence of a phonon environment is investigated
  within a dissipative Hubbard model. For this generalization of the ubiquitous
  spin-boson model the crucial role of symmetries is analysed in detail and
  corresponding invariant subspaces are identified. It is shown that the time
  evolution typically occurs in each of the disjunct subspaces independently
  leading e.g.~asymptotically to a non-Boltzmann equilibrium state. Based on
  these findings explicit results are obtained for two interacting electrons by
  means of a substantially improved real-time quantum Monte Carlo approach. In
  the incoherent regime an appropriate mapping of the many-body dynamics onto
  an isomorphic single particle motion allows for an approximate description of
  the numerical data in terms of rate equations. These results may lead to new
  control schemes of charge transport in tailored quantum systems as
  e.g.~molecular chains or quantum dot arrays.
\end{abstract}


\maketitle
\section{Introduction}
\label{Introduction}

In the last decade enormous progress has been achieved in fabricating designed
quantum systems on the nanoscale. Typical examples include the assembling of
quantum dot arrays to form artificial molecules~\cite{heinzel,doubledot}, the
implementation of synthezised molecular structures in mesoscopic devices to
develop molecular electronics~\cite{molel1,molel2}, and the tailoring of
artificial atoms to obtain fully controllable quantum bits as the basis for
quantum information processing~\cite{quantcomp}. A characteristic feature of
such devices is that they consists of a relatively small number of relevant
degrees of freedom, which are embedded in a much larger surrounding.

Of particular importance are electronic transport processes, which are then
determined by an intimate interplay between electron-electron correlations and
interactions with the environment. In this context, the simplest model,
well-known as the so-called spin-boson model~\cite{dissipativeRMP,weiss},
consists of a two-site chain where a single charge is transferred between the
sites via tunneling hybridization. The charge dynamics can be coherent or
incoherent depending on the interaction with the environmental degrees of
freedom consisting e.g.~of residual vibronic and phonon degrees of freedom or
electromagnetic modes. In the past, the spin-boson model has been proven to
capture essential properties of such electron transfer systems including
various sorts of quantum phase transitions. Together with its extensions to
multi-site systems it describes a broad field of phenomena comprising
e.g.~quantum Brownian motion~\cite{weiss}, Kondo physics~\cite{kondo},
Luttinger liquid behavior~\cite{weiss}, atomic quantum dots~\cite{zwerger}, and
charge transfer in photosynthesis~\cite{jortner}. However, a fundamental
limitation of the spin-boson model is its restriction to single charge
transfer, which becomes particularly severe, when the number of charges and
their dwell time on a certain site can be controlled.

Recently, we studied a generalization to correlated charge transport based on a
dissipative one-dimensional Hubbard model~\cite{prl}. In its isolated form (no
dissipation) the latter one has been extremely well analysed in various
parameter domains and has served as a standard model to reveal the nature of
many-body correlations~\cite{hubbard1}. While those studies have been performed
mostly in the thermodynamic limit to determine equilibrium properties and
quantum phase transitions, the situation in tailored quantum systems is
different: here the dynamics far from equilibrium and in presence of a
dissipative surrounding is to be investigated, where typically only few excess
charges occupy a relatively small to moderate number of sites. One goal lies in
the control of quantum properties in the transport, which inevitably happens to
be influenced by energy exchange with and fluctuations from a phononic
background.

The nonequilibrium dynamics of correlated charge transport in dissipative
environments is an extremely challenging task. This is mainly due to the fact
that on the one hand away from the limit of very weak interaction with the
phonon bath perturbative approaches cannot be applied and on the other hand at
lower temperatures phonon induced retardations become long-ranged. Hence, so
far either completely coherent transfer (no or very weak
dissipation)~\cite{wingreen,ferretti} or completely sequential transfer in the
limit of very strong Coulomb repulsion~\cite{petrov,kohler}, which effectively
reduces to a single charge problem, has been considered.

In the last decade real-time path integral Monte Carlo techniques (PIMC) have
been proven to provide an efficient and numerically exact tool to simulate the
time evolution of dissipative tight-binding systems in ranges where
perturbative schemes are prohibitive~\cite{egger,weiss}. The method has been
substantially improved in the last years to tackle also larger systems (a
larger number of sites) and many charges~\cite{lothar1,lothar2}. These
simulations are challenging in that the dimensionality of the Hilbert space to
be sampled strongly increases with the number of particles, which in turn
aggravates the so-called dynamical sign problem~\cite{suzuki}. The latter one
originates from the oscillatory character of the time evolution operators
generating ''signals'' due to interferences. We note in passing that imaginary
time path integral Monte Carlo approaches have already been successfully
applied to extract thermodynamic properties of the Hubbard
model~\cite{hubbardqmc1,hubbardqmc2,hubbardqmc3}. An alternative approach has
been developed very recently by means of numerical renormalization group
techniques (NRG) and its time dependent generalizations~\cite{tornow}, which
must be seen as being complementary to the PIMC scheme as the former ones are
particularly adapted to the regime of very low temperatures, while the latter
one shows its best efficiency in intermediate temperature domains.

The numerical results in Ref.~\cite{prl} elucidated the crucial relevance of
symmetries of the full Hamiltonian comprising system, bath, and interaction.
Namely, it has been shown that the generic dipole-dipole coupling between
many-electron system and bath leads to the existence of a decoherence free
subspace (DFS), i.e.~a subspace of the total Hilbert space which is invariant
under the time evolution. Interestingly, similar kinds of DFSs have been
analysed in the context of quantum information processing \cite{quantcomp},
however, in many-body systems their structure and impact on the
\textit{transport dynamics} is quite different. For instance, in contrast to
the naive expectation for longer times the dissipative population dynamics of
the many-body states approaches a steady state which is \textit{not} of
Boltzmann type and sensitively depends on the initial preparation. In this
paper we give a more elaborate account of this work and provide a complete
analysis of the symmetry properties of dissipative one-dimensional Hubbard
chains. In fact, a DFS turns out to be just a special kind of an invariant
subspace embedded into a whole class of invariant subspaces. This
characterization provides a detailed understanding of the relaxation towards a
non-Boltzmann type of equilibrium and possibly opens new control mechanisms for
fermionic transport. It further serves in the regime of completely incoherent
transfer as a starting point to approximate the dynamics of the reduced density
matrix by simple master equations with transitions rates obtained from golden
rule calculation, thus generalizing the rate concept known from single charge
transfer \cite{weiss}. Eventually, the existence of invariant subspaces may
substantially improve the efficiency of the PIMC algorithm and particularly
soothe the dynamical sign problem.

The paper is organized as follows. We start in Sec.~\ref{Dissipative
 Hubbard-model} with the description of the dissipative Hubbard model. In the
next two sections, Sec.~\ref{Particle-exchange symmetry} and
Sec.~\ref{Pseudo-angular momentum basis and invariant subspaces}, the
symmetries of the model are analyzed in detail. This provides in
Sec.~\ref{Populations of the localized many-body states} insight into the
dynamical and equilibrium populations, before we give a brief description of
the PIMC method and its description in terms of a rate model in
Sec.~\ref{Quantum Monte-Carlo simulations}. Applications to specific
realizations are discussed in Sec.~\ref{Applications}. At the end are short
summary is given and some conclusions are drawn.

\section{Dissipative Hubbard-model}
\label{Dissipative Hubbard-model}

The system is modeled by an open Hubbard chain with $N$ sites of spacing $q_0$
\begin{eqnarray} \label{H_S}
 H_S &=& \sum_{i=1, \sigma=\uparrow,\downarrow}^N
 \left( E_i \, d^\dag_{i, \sigma}
 \, d^{}_{i, \sigma} + \frac{U_i}{2} \, d^\dag_{i, \sigma} d^{}_{i, \sigma} \,
  d^\dag_{i, -\sigma} d^{}_{i, -\sigma} \right)
\nonumber \\
 && {} + \sum_{i=1, \sigma=\uparrow,\downarrow}^{N-1}
 \Delta_i \left( d^\dag_{i, \sigma} \, d^{}_{i+1, \sigma} +
 h.c. \right) \, ,
\end{eqnarray}
where $d_{i, \sigma}$ and $d^\dag_{i, \sigma}$ are annihilation and creation
operators for electrons, respectively, with spin $\sigma$ on the site $i$.
$E_i$ are the bare energies of the levels, $U_i$ denotes the strengths of the
Coulomb interaction on site $i$, and $\Delta_i$ are the tunneling matrix
elements. The influence of the bosonic bath is given by a Caldeira-Leggett type
of model~\cite{CLM}, where the interaction with the bath
\begin{equation} \label{H_B}
H_B = \sum_\alpha \left(\frac{P_\alpha^2}{2m_\alpha} +
\frac{1}{2}m_\alpha \omega_\alpha^2 X^2_\alpha \right)
\end{equation}
is accomplished via a standard dipole coupling~\cite{jortner}
\begin{eqnarray} \label{H_I}
 H_I = -q_0 \mathcal{P} \sum_\alpha c_\alpha X_\alpha + q_0^2\mathcal{P}^2\sum_\alpha
 \frac{c_\alpha^2}{2m_\alpha \omega_\alpha^2} \, ,
\end{eqnarray}
where
\begin{eqnarray} \label{P}
\mathcal{P} = \sum_{i=1, \sigma}^N \, [i - (N+1)/2]
  \, d_{i, \sigma}^\dag d^{}_{i, \sigma}
\end{eqnarray}
is the polarization operator of the Hubbard chain. This way, the above model
can be seen as a generalization of the spin boson
model~\cite{dissipativeRMP,weiss} of charge transfer to the case of many
electrons and many sites.

In the sequel we deal exclusively with distinguishable charges and thus, in the
case of fermions, concentrate on the two particle/opposite spin sector, see
Ref.~\cite{prl}. We note in passing though that all results presented in this
paper are fully transferable to any two particles as long as they can be
physically distinguished. Based on this analysis, the generalizations to more
charges/particles may be tedious in detail, but straightforward in principle.
The case of indistinguishable particles is related to the bosonic Hubbard model
and will be discussed in a subsequent paper.

A suitable basis for the electronic degrees of freedom is now given in terms of
the localized many-body states (LMBS)
\begin{equation} \label{B_loc}
\mathcal{B}_{\rm loc} =
 \{|s^\downarrow, s^\uparrow\rangle_A\}_{-S \le s^\downarrow, s^\uparrow \le S}
 \,,
\end{equation}
where $|s^\downarrow, s^\uparrow\rangle_A$ denotes an antisymmetrized state
with the $\sigma={\downarrow}$ ($\sigma={\uparrow}$) fermion being localized on
site $i=s^\downarrow+S+1$ ($i=s^\uparrow+S+1$), where $S=(d-1)/2$, i.e.
\begin{equation}
|s^\downarrow, s^\uparrow\rangle_A
 = \frac{1}{\sqrt{2}} \left(|1,\downarrow,s^\downarrow\rangle|2,\uparrow,s^\uparrow\rangle -
 |1,\uparrow,s^\uparrow\rangle|2,\downarrow,s^\downarrow\rangle\right) \,,
\end{equation}
with, e.g., $|1,\downarrow,s^\downarrow\rangle$ denoting a state with the first
particle being in the spin-down state, localized on site $i=s^\downarrow+S+1$.
Since only antisymmetrized states will be used throughout the rest of the
paper, the subscript $A$ will be omitted in the following. The time evolution
of the LMBS populations then follows from
\begin{equation} \label{localized populations}
P_{s^\downarrow, s^\uparrow}(t)
 \equiv \langle s^\downarrow, s^\uparrow|\rho(t)|s^\downarrow,
 s^\uparrow\rangle \,,
\end{equation}
where $\rho(t) = {\rm tr}_B\{W(t)\}$ denotes the reduced density matrix of the
electronic system, obtained from the full density matrix $W(t)=\exp(-iHt/\hbar)
W(0)\exp(iHt/\hbar)$ with $H=H_S+H_B+H_I$ after integrating out the
environmental degrees of freedom (see also Sec.~\ref{Path integral
 representation}).

As shown in~\cite{prl}, the reduced dynamics~(\ref{localized populations}) is
strongly determined by the symmetries of the underlying Hamiltonian of the
total compound. In fact, the dipole type of system-bath coupling provides a
symmetry which gives rise to a decoherence free subspace (DFS). The projector
onto this subspace commutes with $H_S$ and $H_I$, i.e.~with the polarization
$\mathcal{P}$, meaning that it is not affected by the dissipative dynamics of
the environment. Here we argue that such a DFS is actually just a special case
of a more general property: The decomposition of the Hilbert space into
\textit{invariant subspaces} between which neither the free
Hamiltonian~(\ref{H_S}) nor the bath can induce transitions. Accordingly, when
populating initially only one of these subspaces, the dissipative system's
dynamics remains completely restricted to it. This has profound consequences:
(i) Only subspaces of the full Hilbert space must be sampled in the Monte Carlo
simulations which substantially reduces the dynamical sign problem and (ii) for
appropriate bath parameters the full quantum dynamics can be expressed into
simple rate equations, which intrinsically obey the corresponding symmetries
and lead to the correct equilibrium populations.

The analysis of the symmetry properties and its relation to invariant subspace
is most conveniently done by exploiting that the $n$-particle dynamics on a
one-dimensional lattice can be mapped onto an effective single particle
diffusion on a $n$ dimensional one. For instance, as shown in
Ref.~\cite{chemphys}, for two distinguishable fermions on a one-dimensional
lattice with $d$ sites a mapping exists onto one particle dynamics onto a
two-dimensional square lattice with $d^2$ sites, see Fig.~\ref{2d-mapping for
 d=3}. This mapping turns out to be a very convenient tool on the one hand to
visualize the relation between many-body dynamics and dissipation and on the
other it may serve as a starting point to apply perturbative methods as
e.g.~the non-interacting blip/cluster approximation (NIBA/NICA), which are
known to be powerful means to capture the single particle motion. The rate
equations mentioned above are a direct consequence of this type of
approximation.
%
\begin{figure}
\vspace*{0.7cm}
\epsfig{file=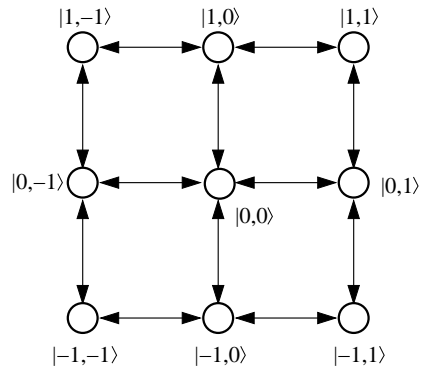, height=4.8cm}
\caption[]{\label{2d-mapping for d=3}
2d mapping of the LMBS~(\ref{B_loc}) for $d=3$.} 
\vspace*{0.0cm}
\end{figure}
%

\section{Spin-permutation symmetry}
\label{Particle-exchange symmetry}

Since the dissipative Hubbard Hamiltonian contains no terms depending on the
fermions' spins, it is invariant with respect to any permutation of the latter.
This leads to the most robust symmetry of the model, namely,
\begin{equation}
[\mathcal{F}, H] = [\mathcal{F}, H_S] = [\mathcal{F}, H_I]
= [\mathcal{F}, H_B] = 0\, ,
\end{equation}
where we introduced the spin-permutation operator
\begin{equation} \label{spin-permuation operator}
\mathcal{F} |s^\downarrow, s^\uparrow\rangle
 = |s^\uparrow, s^\downarrow\rangle \,,
\end{equation}
which, for two particles, is equivalent to a simultaneous spin flip.

Note that $\mathcal{F}$ commutes not just with the total Hamiltonian $H$, but
with $H_0$ and $H_I$ separately, such that we can easily find an electronic
basis of simultaneous eigenstates of $\mathcal{F}$ and $\mathcal{P}$. For
example
\begin{eqnarray} 
\mathcal{B}_{\rm flip} &\equiv& \mathcal{B}^+_{\rm flip} \cup \mathcal{B}^-_{\rm flip} \,; \nonumber\\
\mathcal{B}^+_{\rm flip} &\equiv&
\{|s^\downarrow, s^\uparrow = s^\downarrow\rangle\}_{|s^\downarrow| \le S} \nonumber\\
&& \cup \left\{\frac{1}{\sqrt{2}} \left(|s^\downarrow,
s^\uparrow\rangle +
|s^\uparrow,s^\downarrow\rangle\right)\right\}_{s^\downarrow
< s^\uparrow}
 \,, \nonumber\\
\mathcal{B}^-_{\rm flip} &\equiv&
\left\{\frac{1}{\sqrt{2}} \left(|s^\downarrow,
s^\uparrow\rangle -
|s^\uparrow,s^\downarrow\rangle\right)\right\}_{s^\downarrow
< s^\uparrow} \,,
\end{eqnarray}
where $\mathcal{B}^\pm_{\rm flip}$ only includes eigenstates of $\mathcal{F}$
with eigenvalue $\pm 1$, respectively. Transitions between $\mathcal{B}^+_{\rm
 flip}$ and $\mathcal{B}^-_{\rm flip}$ are globally prohibited,
\textit{independent of the system parameters $d$, $E_i$, $U_i$, $\Delta_i$, or
 any bath parameters}. Therefore, the electronic Hilbert space $\mathcal{H}$
decomposes into two \textit{disjunctive invariant subspaces} $\mathcal{G}^\pm
\equiv {\rm span}\{\mathcal{B}^\pm_{\rm flip}\}$, i.e.
\begin{equation}
\mathcal{H} = \mathcal{G}^+ \cup \mathcal{G}^- \,,\quad
\mathcal{G}^+ \cap \mathcal{G}^- = \O \,,\quad
H\mathcal{G}^+ \subseteq \mathcal{G}^+ \,,~ H\mathcal{G}^-
\subseteq \mathcal{G}^- \,,
\end{equation}
with distinct routes to thermal equilibrium. As a consequence, the LMBS
populations~(\ref{localized populations}) exhibit different relaxation time
scales and, even at high temperatures, a non-Boltzmann equilibrium
distribution. Furthermore, the latter explicitly depends on the \textit{initial
 preparation}. For example, for an energetically completely degenerate system
(i.e.~$E_i = U_i = 0$) the populations for the $d(d+1)/2$ states in
$\mathcal{B}^+_{\rm flip}$ equilibrate towards $2P^+_0/d(d+1)$, while those for
the $d(d-1)/2$ states in $\mathcal{B}^+_{\rm flip}$ equilibrate towards
$2P^-_0/d(d-1)$, where $P^\pm_0$ denote the initial populations of the
subspaces $\mathcal{G}^\pm$, respectively ($P^+_0 + P^-_0 = 1$). Accordingly,
even for sufficiently strong damping and/or high temperatures, the equilibrium
populations of the localized states~(\ref{B_loc}) do \textit{not} coincide with
a Boltzmann distribution.

For the special case $d=2$, it is straightforward to show that the basic
spin-permutation symmetry fully explains the existence of a decoherence-free
state as reported in Ref.~\cite{prl} (namely in form of the only element of
$\mathcal{B}^-_{\rm flip}$). The robustness of this DFS with respect to changes
in system and bath parameters directly follows from the robustness of the
spin-permutation symmetry itself. For $d>2$ we postpone a more detailed
discussion and first analyse in the next section additional symmetries related
to particular parameter ranges for $E_i, U_i, \Delta_i$. We already note,
though, that the spin-permutation symmetry can be destroyed e.g.~by applying
external magnetic fields, which in turn would also allow to control its
influence on the many-body dynamics.

\section{Pseudo-angular momentum basis and invariant subspaces}
\label{Pseudo-angular momentum basis and invariant subspaces}

For the tight binding lattice with the localized basis~(\ref{B_loc}), it is
very suggestive to identify the discrete positions of the charges
$s^\downarrow$ and $s^\uparrow$ with the discrete eigenvalues of the
z-components $J^\downarrow_z$ and $J^\uparrow_z$, respectively, of two
pseudo-angular momentum operators. Accordingly, the latter ones obey
$\vec{J}^{\downarrow2} = \vec{J}^{\uparrow2} = \hbar^2 S(S+1)$ with $d=2S+1$.
For single charge transfer on two sites this leads to the famous spin-boson
model where the polarization is then proportional to the Pauli matrix
$\sigma_z$. Now, for two charges the polarization $\mathcal{P}$ is proportional
to $J^\downarrow_z+J^\uparrow_z$. Hence, for the analysis of symmetry
properties it is convenient to introduce an alternative basis of the fermionic
Hilbert space following from a \textit{total} pseudo-angular momentum $\vec{J}
= \vec{J}^\downarrow + \vec{J}^\uparrow$. The consequences of this formulation
are studied in this section.

\subsection{Collective states}
\label{Collective states}

Within the pseudo-angular momentum representation the localized states in
$\mathcal{B}_{\rm loc}$~(\ref{B_loc}) are simultaneous eigenstates of
$\vec{J}^{\downarrow2}$, $J^\downarrow_z$, $\vec{J}^{\uparrow2}$, and
$J^\uparrow_z$ due to
\begin{equation} 
|s^\downarrow, s^\uparrow\rangle \equiv |j^\downarrow=S,
j^\uparrow=S; m^\downarrow=s^\downarrow,
m_\uparrow=s^\uparrow\rangle \,,
\end{equation}
while, as already mentioned, the polarisation operator~(\ref{P}) can be
expressed as
\begin{equation} \label{polarisation}
 \mathcal{P} = \frac{J_z^\uparrow + J_z^\downarrow}{\hbar}
 = \frac{J_z}{\hbar} \,. 
\end{equation}
Here $J_z$ denotes the $z$ component of $\vec{J} = \vec{J}^\downarrow +
\vec{J}^\uparrow$. An alternative electronic basis set is then given by the
simultaneous eigenstates of these two operators, i.e.,
\begin{equation} \label{B_J}
\mathcal{B}_{\rm J} = \{|\psi^{(j)}_m\rangle\}_{0 \le j \le
d-1, -j \le m \le j} \, ,
\end{equation}
where $\vec{J}^2 |\psi^{(j)}_m\rangle = \hbar^2 j(j+1) |\psi^{(j)}_m\rangle$
and $J_z |\psi^{(j)}_m\rangle = \hbar m |\psi^{(j)}_m\rangle$. Note that, like
the localized states in $\mathcal{B}_{\rm loc}$, the collective states in
$\mathcal{B}_{\rm J}$ are also eigenstates of the polarization operator
$\mathcal{P}$~(\ref{polarisation}), as well as of the spin-permutation operator
$\mathcal{F}$~(\ref{spin-permuation operator}) ($\mathcal{F}
|\psi^{(j)}_m\rangle = (-1)^{j+1} |\psi^{(j)}_m\rangle$).

The transformation between the two basis sets is conveniently given in terms of
the (real-valued) Clebsch-Gordan coefficients~\cite{clebsch} $\langle
s^\downarrow, s^\uparrow|\psi^{(j)}_m\rangle \equiv \langle S,S,s^\downarrow,
s^\uparrow|\psi^{(j)}_m\rangle$. Hence, defining
\begin{equation} 
C^{(j)}_{s^\downarrow, s^\uparrow}
 \equiv \langle s^\downarrow, s^\uparrow|\psi^{(j)}_{s^\downarrow
      + s^\uparrow}\rangle
\end{equation}
and exploiting $\langle s^\downarrow, s^\uparrow|\psi^{(j)}_m\rangle = 0$ for
$m \ne s^\downarrow + s^\uparrow$, we obtain
\begin{eqnarray} \label{basis transformation}
|\psi^{(j)}_m\rangle &=& \sum_{s^\downarrow =
\max\{-S,m-S\}}^{\min\{S,m+S\}}
           C^{(j)}_{s^\downarrow, m-s^\downarrow}
           |s^\downarrow,m-s^\downarrow\rangle \,, \nonumber\\
|s^\downarrow, s^\uparrow\rangle &=& \sum_{j=|s^\downarrow
+ s^\uparrow|}^{d-1} C^{(j)}_{s^\downarrow, s^\uparrow}
|\psi^{(j)}_{s^\downarrow+s^\uparrow}\rangle \,. 
\end{eqnarray}
With the aid of Eq.~\ref{basis transformation} one can now easily express the
LMBS populations~(\ref{localized populations}) in terms of the corresponding
ones in the pseudo-angular momentum basis,
\begin{equation} \label{angular momentum populations}
\hat{P}_{j,m}(t) \equiv
\langle\psi^{(j)}_m|\rho(t)|\psi^{(j)}_m\rangle \,,
\end{equation}
yielding
\begin{eqnarray} \label{LMBS populations}
\lefteqn{P_{s^\downarrow,s^\uparrow}(t) =
\sum_{j=|s^\downarrow+s^\uparrow|}^{d-1}
C^{{(j)}^{\scriptstyle 2}}_{s^\downarrow,s^\uparrow}
\hat{P}_{j,s^\downarrow+s^\uparrow}(t)} \nonumber\\
&&{} + 2\sum_{\scriptstyle j,j'=|s^\downarrow+s^\uparrow|
\atop \scriptstyle j'>j}^{d-1}
  C^{(j)}_{s^\downarrow,s^\uparrow} C^{(j')}_{s^\downarrow,s^\uparrow}
  {\rm Re}\{\langle\psi^{(j')}_{s^\downarrow+s^\uparrow}
  |\rho(t)|\psi^{(j)}_{s^\downarrow+s^\uparrow}\rangle\} \nonumber\\
\end{eqnarray}
Equation~(\ref{LMBS populations}) also governs the transformation of the
initial populations between the two basis sets. Particularly simple expressions
are gained for an initially factorizing system-bath state where, as in many
experiments, the system is prepared to reside on some localized state
$|s^\downarrow_0,s^\uparrow_0\rangle$. The corresponding initial populations
$\hat{P}_{j,m}(0)$ are then obtained from
\begin{equation} \label{P_0}
\hat{P}_{j,m}(0) = \left\{
\begin{array}{cl}
C^{{(j)}^{\scriptstyle 2}}_{s^\downarrow_0,s^\uparrow_0}
 & {\rm if}\ m=s^\uparrow_0+s^\downarrow_0 \,,\ j \ge |s^\uparrow_0+s^\downarrow_0| \,, \\
0 & {\rm else} \,. 
\end{array}
\right. 
\end{equation}

\subsection{Invariant subspaces}
\label{Invariant subspaces}

The formulation based on the total pseudo-angular momentum now allows to
identify a further symmetry of the dissipative Hubbard model. For this purpose
we first consider the case, where the full Hamiltonian $H$ commutes with the
total pseudo-angular momentum $\vec{J}^{\, 2}$, thus leading to invariant
subspaces similar to the symmetry noted in Sec.~\ref{Particle-exchange
 symmetry}. Classically speaking, $H$ preserves the angle between the
individual pseudo-angular momenta $J^\downarrow$ and $J^\uparrow$.

Accordingly, we look for a representation of the electronic Hamiltonian $H_S$
in terms of the components $J_x$, $J_y$, $J_z$. Turning first to the
nearest-neighbor hopping term $H^\Delta_S \equiv \sum_{i,\sigma} \Delta_i (
d^\dag_{i, \sigma} \, d^{}_{i+1, \sigma} + h.c.)$, it is straightforward to
show that with respect to $\mathcal{B}_{\rm loc}$, it exhibits the same
structure as $J_x$, namely,
\begin{equation} 
\langle\hat{s}^\downarrow,\hat{s}^\uparrow|H^\Delta_S|s^\downarrow,s^\uparrow\rangle
= 0 \quad \Leftrightarrow \quad
\langle\hat{s}^\downarrow,\hat{s}^\uparrow|J_x|s^\downarrow,s^\uparrow\rangle
= 0 \,,
\end{equation}
provided all couplings $\Delta_i$ are non-zero. 

Since the non-vanishing matrix elements of $J_x$ are given by
\begin{equation} 
\langle
s^\downarrow\pm1,s^\uparrow|J_x|s^\downarrow,s^\uparrow\rangle
= \frac{\hbar}{2} \sqrt{S(S+1) - s^\downarrow(s^\downarrow
\pm 1)}
\end{equation}
(and $\langle s^\downarrow,s^\uparrow\pm1|J_x|s^\downarrow,s^\uparrow\rangle$
accordingly), one finds that $H^\Delta_S\propto J_x$ is always true for $d=2$.
For $d>2$ this holds only if the tunnel couplings (hopping amplitudes) are
chosen properly, e.g.~for $d=3$ one must have $\Delta_1 = \Delta_2$,
i.e.~isotropic coupling, and $d \ge 4$ requires quite specific values for the
coupling strengths. We will discuss this point in more detail below.

Next, we turn to the on-site terms in $H_S$ comprising the on-site energies and
the Coulomb interactions. Since the latter one effectively gives a contribution
to the former ones, we find that if $E_i$ and $U_i$, $i=1, \ldots N$, are
distributed up to an overall factor according to the matrix elements of $J_z$,
i.e.,
\begin{equation}
\langle\hat{s}^\downarrow,\hat{s}^\uparrow|J_z|s^\downarrow,s^\uparrow\rangle
= (s^\downarrow+s^\uparrow)
\delta_{\hat{s}^\downarrow=s^\downarrow}\delta_{\hat{s}^\uparrow=s^\uparrow}
\, ,
\end{equation}
then the electronic Hamiltonoperator can be cast into the form
\begin{equation} \label{pseudo-spin H_S}
H_S = E J_z + \hat{\Delta} J_x \,,
\end{equation}
with appropriate constants $E$ and $\hat{\Delta}$. It then follows immediately
that
\begin{equation} \label{J^2 commutator}
[H, \vec{J}^{\, 2}] = [H_S, \vec{J}^{\, 2}] = [H_I,
\vec{J}^{\, 2}] = [H_B, \vec{J}^{\, 2}] = 0 \,. 
\end{equation}
The total pseudo-angular momentum is thus a conserved quantity with respect to
the \textit{full dissipative system}. Equivalently, for a dissipative Hubbard
model with $H_S$ of the form~(\ref{pseudo-spin H_S}) both the free system as
well as the bath can induce only transitions between states with the same total
pseudo-angular momentum. Of course, Eq.~(\ref{pseudo-spin H_S}) is a
\textit{sufficient but not necessary} condition for Eq.~(\ref{J^2 commutator})
to hold. Obviously, any $H_S$ and $H_I$ where the system part can be written in
terms of $J_x$, $J_y$, and $J_z$ (and powers of these operators) fulfill
Eq.~(\ref{J^2 commutator}); in particular, the coupling between system and bath
may also be nonlinear in the system operator. Indeed, below we will present an
example where Eq.~(\ref{pseudo-spin H_S}) does not hold, while the latter
relation still applies.

With respect to the reduced electronic dynamics, Eq.~(\ref{J^2 commutator})
leads to consequences similar to those described in Sec.~\ref{Particle-exchange
 symmetry}. First, the electronic subspace $\mathcal{H}$ of the full Hilbert
space can be further decomposed into \textit{disjunctive invariant} subspaces,
\begin{equation} \label{decomposition}
\mathcal{H} = \cup_{j=0}^{d-1} \mathcal{G}_j \,,
\end{equation}
where $\mathcal{G}_j$ denotes the subspace spanned by all eigenstates of
$\vec{J}^2$ with a fixed total pseudo-spin $j$,
\begin{equation} 
\mathcal{G}_j = {\rm span}\{|\psi^{(j)}_{-j}\rangle,
\dots, |\psi^{(j)}_j\rangle\} \,,
\end{equation}
with the properties
\begin{equation} \label{invariance}
\mathcal{G}_k \cap \mathcal{G}_j = \o \quad {\rm for\
all}\ k \ne j \,, \quad H\mathcal{G}_j \subseteq
\mathcal{G}_j \,. 
\end{equation}
To put it differently, with the above decomposition we have identified the
irreducible representations of the sum of two angular momentum operators each
with angular momentum $S=(d-1)/2$. Unlike the spin-permutation symmetry, now
the number of invariant subspaces equals $d$, thus explicitly depending on the
length of the tight binding lattice. Furthermore, the structure of these
subspaces turns out to resemble \textit{linear chains} since due to
\begin{eqnarray} 
\lefteqn{\langle\psi^{(j)}_{m'}|H_S|\psi^{(j)}_m\rangle
= E\delta_{m'=m}} \nonumber\\
&&{} + \frac{\hat{\Delta}}{2} \left(\sqrt{j(j+1) - m(m+1)}
    \delta_{m'=m+1} \right. \nonumber\\
&&\qquad{}\left.  + \sqrt{j(j+1) -
m(m-1)}\delta_{m'=m-1}\right)\, . 
\end{eqnarray}
Transitions thus are only possible between nearest neighbors with respect to
$m$ between states $|\psi^{(j)}_m\rangle$ with fixed $j$.

\begin{figure}
\vspace*{0.7cm}
\epsfig{file=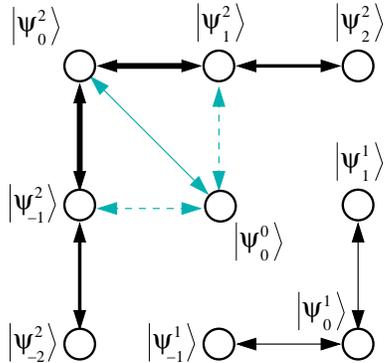, height=4.8cm}
\caption[]{\label{d_3_invariant_subspaces} 2d mapping for $d=3$ and the angular-momentum
  states~(\ref{B_J}). Black lines indicate the couplings for $\epsilon_i =
  U = 0$; the thickness of the lines is proportional to the square of the
  coupling strengths. Gray solid lines indicate additional couplings for $U \ne
  0$, $\epsilon_i \ne 0$; gray dashed lines indicate additional couplings for
  anisotrope couplings $\Delta_1 \ne \Delta_2$.} 
\vspace*{0.0cm}
\end{figure}

To summarize the main outcome of this analysis, we have shown that an
electronic Hamiltonian obeying~(\ref{pseudo-spin H_S}) leads to a Hilbert space
basis which can be decomposed into one singlet part, one triplet part, one
quintet part, \ldots, up to a $2(d-1)+1$ part, each of which has a linear
structure and between which transitions induced by the Hamiltonian are not
possible whatsoever (see Fig.~\ref{d_3_invariant_subspaces}). The subspaces
solely interact via the coupling to a common bath. The singlet space is
exceptional in that it is an eigenstate of $H_I$ with vanishing eigenvalue for
the polarization operator, thus being protected by any dynamical time evolution
of the dissipative system, i.e.~a DFS. 

%

\section{Populations of the localized many-body states}
\label{Populations of the localized many-body states}

For transport processes the populations of the LMBS are of particular
relevance. Here, we show how the existence of the ideal symmetry~(\ref{J^2
 commutator}) may simplify their calculation considerably. This gives also
direct access to the stationary equilibrium populations which are approached
for sufficiently long times.

\subsection{Symmetries and population dynamics}
\label{Symmetries and population dynamics}

Due to the disjunct structure of the full Hilbert space, see
Eq.~(\ref{invariance}), not only the \textit{total} population is conserved,
but also the \textit{individual} ones within each invariant subspace. In
particular, for a factorizing initial preparation $W(0)=\rho_0 \exp(-\beta
H_B)$, where $\rho_0=|\varphi_0\rangle\langle|\varphi_0|$ presents a projector
on some pure fermionic state $|\varphi_0\rangle$, subspaces with no overlap
with $|\varphi_0\rangle$ do not participate in the dynamics at all. This leads
to a reduction of the dimensionality of the relevant Hilbert space and thus of
the complexity of the system to be treated numerically. If the initial
electronic population is even restricted to a single subspace,
i.e.~$|\varphi_0\rangle \in \mathcal{G}_j$, the corresponding dynamics is
identical to that of a single particle moving on a linear chain with just
$2j+1$ sites. For $|\varphi_0\rangle \in \mathcal{B}_{\rm loc}$, this scenario
only applies to the states $|{-S},-S\rangle$
(i.e.~$|\psi^{(d-1)}_{-(d-1)}\rangle$) and $|S,S\rangle$
(i.e.~$|\psi^{(d-1)}_{d-1}\rangle$) with $d=j+1$. We note in passing that this
property also applies to the situation $|\varphi_0\rangle \in \mathcal{G}_j
\cup \mathcal{G}_0$ since $\mathcal{G}_0$ does not exhibit any dynamical
evolution at all. For other initial preparations within $\mathcal{B}_{\rm
 loc}$, however, direct phonon induced couplings between the participating
subspaces mediated by the common environment~\cite{chemphys} have to be taken
into account, which do influence the dynamics but conserve the population in
each subspace. For initial states which have overlaps with all invariant
subspaces (i.e.~$|{-S},S\rangle$ or $|S,-S\rangle$), the dynamics only takes
places in a $(d^2-1)$ dimensional subspace since again $\mathcal{G}_0$ does not
participate. 

Due to the invariance of the subspaces $\mathcal{G}_j$ with respect to $H$, the
cross terms $\langle\psi^{(j')}_{s^\downarrow+s^\uparrow}
|\rho(t)|\psi^{(j)}_{s^\downarrow+s^\uparrow}\rangle$ in Eq.~(\ref{LMBS
 populations}) are only non-vanishing if the initial state $|\varphi_0\rangle$
has an overlap with both $\mathcal{G}_j$ and $\mathcal{G}_{j'}$. For most
preparations in a LMBS, this significantly simplifies Eq.~(\ref{LMBS
 populations}). This holds especially for preparations where both fermions
enter at the same end of the chain (i.e.~$|{-S},-S\rangle$ and $|{S},S\rangle$,
both $\in \mathcal{G}_{d-1}$), when all localized populations can be expressed
in terms of the pseudo-angular momentum populations alone. For all other
localized preparations, however, the cross terms play a crucial role in the
short-to-intermediate time domain. On the other hand, using Eq.~(\ref{LMBS
 populations}) to obtain information about these cross terms also eludicates
the dynamical evolution of coherences in the system.

\subsection{Equilibrium populations}
\label{Equilibrium populations}

For long times a dissipative system reaches a stationary state, typically a
thermal equilibrium where all states are occupied according to a Boltzmann
distribution. In the case studied here, namely, a dissipative Hubbard model
with the symmetry relations~(\ref{J^2 commutator}), the same applies not to the
total Hilbert space, but rather its disjunct subspaces. In particular, for a
state $|\psi^{(j)}_m\rangle \in \mathcal{G}_j$ one obtains
\begin{equation} \label{equil. pops B_J}
\hat{P}^\infty_{j,m}\equiv\lim_{t\to
\infty}\hat{P}_{j,m}(t)
 = \hat{P}^{(j)}_0 \frac{e^{-\hbar \beta \hat{\epsilon}^{(j)}_m}}
   {\sum_{m'=-j}^j e^{-\hbar \beta \hat{\epsilon}^{(j)}_{m'}}} \,,
\end{equation}
where the initial population in $\mathcal{G}_j$ is $\hat{P}^{(j)}_0 \equiv
\sum_{m=-j}^j \hat{P}_{j,m}(0)$, and $\hbar \hat{\epsilon}^{(j)}_m \equiv
\langle\psi^{(j)}_m|H_S|\psi^{(j)}_m\rangle$. Accordingly, the final
occupations of the states $|\psi^{(j)}_m\rangle$ do not obey a Boltzmann
distribution, i.e.,
\begin{equation} 
\frac{\hat{P}^\infty_{j,m}}{\hat{P}^\infty_{j',m'}} \ne
e^{-\hbar \beta(\hat{\epsilon}^{(j)}_m -
\hat{\epsilon}^{(j')}_{m'})} \,. 
\end{equation}
The same is true for the states in $\mathcal{B}_{\rm loc}$. Since in thermal
equilibrium all cross terms in Eq.~(\ref{LMBS populations}) vanish
($P^\infty_{s^\uparrow,s^\downarrow} = P^\infty_{s^\downarrow,s^\uparrow}$), we
have
\begin{equation}
\label{LMBS equil. pops} P^\infty_{s^\downarrow,s^\uparrow}
= \sum_{j=|s^\downarrow+s^\uparrow|}^{d-1}
  C^{{(j)}^{\scriptstyle 2}}_{s^\downarrow,s^\uparrow} \frac{\hat{P}^{(j)}_0}{Z_j}
  e^{-\hbar\beta\hat{\epsilon}^{(j)}_{s^\downarrow+s^\uparrow}}
\end{equation}
with $Z_j = \sum_{m'=-j}^j e^{-\hbar \beta \hat{\epsilon}^{(j)}_{m'}}$.
Apparently, the equilibrium populations strongly depend on the initial
preparation, i.e.~on $\hat{P}^{(j)}_0$.

\section{Quantum Monte-Carlo simulations}
\label{Quantum Monte-Carlo simulations}

So far we have focused on a convenient representation of the states within the
fermionic subspace of the total Hilbert space. To obtain the populations
Eqs.~(\ref{localized populations}) and~(\ref{angular momentum populations})
explicitly, however, one must find an adequat treatment of the environmental
degrees of freedom as well. The path integral approach has been proven to allow
for an exact elimination of the latter and to provide in combination with Monte
Carlo techniques a numerically exact evaluation.

\subsection{Path integral representation}
\label{Path integral representation}

The expectation value of a system observable $A$
\begin{equation}
\langle A(t)\rangle = \mbox{Tr}_{\rm S}\{\rho(t) A\}
\end{equation}
requires the knowledge of the reduced density operator
\begin{equation}
\rho(t)=\mbox{Tr}_{\rm B}\left\{{\rm e}^{-iHt/\hbar} W(0)
{\rm e}^{iHt/\hbar}\right\}\, . 
\end{equation}
Here, $H$ denotes the full Hamiltonian~(\ref{H_S})--(\ref{H_I}), $W(0)$ is an
initial density, and the trace is performed over the bath degrees of freedom
only. The details of the path integral representation for $\rho(t)$ have been
given elsewhere~\cite{dissipativeRMP,weiss,egger,lothar1}. Here, the standard
expression known e.g.~for the spin-boson model has to be extended to capture
the dynamics of two distinguishable fermions. Accordingly, the reduced density
operator is expressed as a double path integral along a Keldysh contour with
forward $s^\sigma$ and backward $\tilde{s}^\sigma$ paths corresponding to the
basis set $\mathcal{B}_{\rm loc}$. The impact of the dissipative environment
appears as an influence functional introducing arbitrarily long-ranged
interactions in time along and between the paths. It is convenient to switch to
the sum and difference coordinates $\eta^\sigma = s^\sigma + \tilde{s}^\sigma$
and $\xi^\sigma = s^\sigma - \tilde{s}^\sigma$, respectively, so that one
arrives for the expectation value at the exact expression
\begin{equation} \label{path integral expression for A(t)}
\langle A(t)\rangle = \oint\!\mathcal{D}\vec{\eta}
\oint\!\mathcal{D}\vec{\xi}\; a[\vec{\eta},\vec{\xi}]\,
\mathcal{K}[\vec{\eta},\vec{\xi}]\, \exp\!\left\{ -
 \Phi[\vec{\eta},\vec{\xi}\,] \right\} \; ,
\end{equation}
where $a[\vec{\eta},\vec{\xi}]$ is the measurement functional corresponding to
$A$ in terms of the combined system paths $\vec{\eta}(t) = (\eta^\uparrow(t),
\eta^\downarrow(t))$ and $\vec{\xi}(t) = (\xi^\uparrow(t), \xi^\downarrow(t))$.
Furthermore, $\mathcal{K}$ is the bare action factor in absence of a reservoir,
and the influence functional reads
\begin{eqnarray} \label{influence exponent}
\Phi[\vec{\eta},\vec{\xi}] &=& \int_0^t ds \int_0^s du
\left\{\left[\vec{\xi}(s) \cdot \vec{e}\right] L'(s-u)
\left[\vec{\xi}(u) \cdot \vec{e}\right]\right. \nonumber\\
&&\left.\hspace{1.5cm}{} + i \left[\vec{\xi}(s) \cdot
\vec{e}\right] L''(s-u)
\Big[\vec{\eta}(u) \cdot \vec{e}\Big]\right\} \nonumber\\
&&{} +i \frac{\mu}{2}\int_0^t ds \left[ \vec{\xi}(s) \cdot
\vec{e}\right]\Big[\vec{\eta}(s) \cdot \vec{e}\Big]
\end{eqnarray}
with $\vec{e} = (1, 1)$. The kernel $L(t) = L'(t) + i L''(t)$ is related to the
force-force auto-correlation function of the bath and is completely determined
by the spectral density of its modes~(\ref{L(t)}). Further, $\mu =
\lim_{\hbar\beta \to 0} \hbar\beta L(0)$. Note that even in absence of Coulomb
interaction the two charges are correlated due to the coupling to the common
heat bath.

An analytical treatment of the expression~(\ref{path integral expression for
 A(t)}) is in general not feasible, mainly due to the retardations in the
influence functional which grow with decreasing temperature, roughly as
$\hbar\beta$. In this situation PIMC methods have been shown to be very
powerful and numerically exact means to explore the non-perturbative range.
Appendix~\ref{Appendix} explains in some detail our approach which is
straightforwardly obtained from the formally very similar case of a single
dissipative particle~\cite{dissipativeRMP,weiss,egger,lothar1} and which can be
easily extended to the cases of more than two or indistinguishable
particles~\cite{chemphys}.

A striking difference to the single-particle case, however, is the existence of
symmetries as described above and the consequent decomposition of the full
system's Hilbert state into invariant subspaces. To exploit these symmetries
for our numerical studies, we express Eq.~(\ref{path integral expression for
 A(t)}) in terms of system paths $J(t)$ and $M_J(t)$, referring to the states
$|\psi_M^{(J)}\rangle$~(\ref{B_J}) rather than in terms of $s^\downarrow(t)$
and $s^\uparrow(t)$. It turns out that the environmental influence can again be
summarized in an influence functional which is obtained from
Eq.~(\ref{influence exponent}) by simply replacing $\vec{\eta}(s) \cdot
\vec{e}$ and $\vec{\xi}(s) \cdot \vec{e}$ with $\eta_M \equiv M_J(t) +
\tilde{M}_J(t)$ and $\xi_M \equiv M_J(t) - \tilde{M}_J(t)$, respectively, where
$M_J(t)$ and $\tilde{M}_J(t)$ again denote paths on the forward and backward
part of the Keldysh contour, respectively. The corresponding influence
functional then reads
\begin{eqnarray} \label{influence exponent for M_J}
\Phi[\eta_M,\xi_M] &=& \int_0^t ds \int_0^s du
\left[\xi_M(s) L'(s-u) \xi_M(u)\right. \nonumber\\
&&\left.\hspace{1.5cm}{} + i \, \xi_M(s) L''(s-u)
\eta_M(u)\right] \nonumber\\
&&{} +i \frac{\mu}{2}\int_0^t ds\, \xi_M(s) \eta_M(s) \,. 
\end{eqnarray}
Note that for fixed $J$ this is just the influence functional of a single
dissipative particle residing on $2J+1$ discrete states.

Now, if Eq.~(\ref{J^2 commutator}) holds, $J(t)$ is conserved upon propagation
with $\exp(\pm i t H/\hbar)$ (see Eq.~(\ref{discretization})). A change in
$J(t)$ can thus only be mediated by the initial preparation Tr$\{W(0)\}$ and
the measurement $a[\eta_M,\xi_M]$ at time $t$; otherwise $J(t)$ remains
constant. In terms of paths of $J$ and $M$, Eq.~(\ref{path integral expression
 for A(t)}) therefore eventually becomes
\begin{eqnarray} \label{decomposed path integral expression for A(t)}
\langle A(t)\rangle &=& \sum_{J_1,J_2=0}^{d-1}
A_{J_1,J_2}(t) \,,
\nonumber\\
A_{J_1,J_2}(t) &=&
\oint\!\mathcal{D}\eta_M\oint\!\mathcal{D}\xi_M\;
 a_{J_1,J_2}[\eta_M,\xi_M]
 \nonumber\\
&&\times \mathcal{K}_{J_1,J_2}[\eta_M,\xi_M] \exp\!\left\{
-\Phi[\eta_M,\xi_M] \right\} ,
\end{eqnarray}
where $J_1$ and $J_2$ denote the values of the piecewise constant $J$ paths
along the forward and backward parts of the Keldysh contour, respectively, and
$a_{J_1,J_2}[\eta_M,\xi_M]$ is the measurement functional of the operator
$\mathbb{P}_{J_1}A\mathbb{P}_{J_2}$, with the projector $\mathbb{P}_j$ onto the
subspace $\mathcal{G}_j$.

Further simplification can be achieved by exploiting that $A_{J_1,J_2}(t)
\equiv 0$ if any of the subspaces $\mbox{Tr}_{\rm B}\{
W(0)\}\mathcal{G}_{J_1}$, $ \mbox{Tr}_{\rm B}\{ W(0)\}\mathcal{G}_{J_2}$ or
$A\mathcal{G}_{J_1}$, $ A\mathcal{G}_{J_2}$ are empty. For example, if the
system is initially prepared in the localized state $|{-S},-S\rangle$, one
simply obtains
\begin{equation} 
\langle A(t)\rangle = A_{d-1,d-1}(t) \,. 
\end{equation}
Accordingly, in cases where the initial preparation does not impose transitions
between subspaces with different $J$, the dissipative two-particle system can
be decomposed into \textit{d independent dissipative single-particle systems}.
For PIMC simulations this presents a major progress since each of the systems
representing the $A_{J_1,J_2}(t)$ contributions exhibits a significantly lower
dimensionality than the full $d^2$ dimensional two-particle system. Hence, by
evaluating Eq.~(\ref{decomposed path integral expression for A(t)}) rather than
Eq.~(\ref{path integral expression for A(t)}), the \textit{dynamical sign
 problem}~\cite{suzuki}, which stems from interferences between different
quantum paths and grows exponentially with the size of the system, can be
greatly reduced. In the quantum regime this allows for substantial
computational savings, despite the fact that the evaluation of
Eq.~(\ref{decomposed path integral expression for A(t)}) requires $d(d-1)/2$
independent PIMC simulations (note $A_{J_2,J_1}(t) = A^\ast_{J_1,J_2}(t)$).

In case that Eq.~(\ref{J^2 commutator}) does not apply, an expression similar
to Eq.~(\ref{decomposed path integral expression for A(t)}) can still be
obtained exploiting the spin-permutation symmetry. Therefore, the arguments
presented above concerning the computational costs of PIMC simulations still
apply, albeit to a lower degree.

For further details of the approach and particularly of the strategy to soothe
the dynamical sign problem we refer to the
literature~\cite{egger,lothar1,lothar2}. The basic ingredients are this: (i)
One exploits the fact that the influence functional depends only linearly on
the quasi-classical paths $\vec{\eta}(t)$, so that in Eq.~(\ref{path integral
 expression for A(t)}) the corresponding summations can be expressed as a
series of simple matrix multiplications and therefore be carried out
explicitly; (ii) for the sampling procedure one chooses an MC weight with no
long-time retardations. Since they are fully taken into account in the final
accumulation process evaluating the exact expression~(\ref{path integral
 expression for A(t)}), the numerical exactness of the MC scheme is not
impaired. This way, one achieves a strong decoupling of quasi-classical
($\vec{\eta}$) and quantum ($\vec{\xi}$) coordinates, which allows to store
products of short time propagators independent of the MC-sampling. Eventually a
speed-up with respect to the original method~\cite{egger} by a factor of about
100 is gained. Of course, this strategy can be applied for evaluating both
Eq.~(\ref{path integral expression for A(t)}) as well as Eq.~(\ref{decomposed
 path integral expression for A(t)}).

\subsection{Rate model}
\label{Rate model}

For one-dimensional tight-binding systems a description of the single particle
population dynamics in terms of simple master equations with transition rates
gained from golden rule calculations is known to be quite accurate for
sufficiently strong dissipation and/or high temperatures (incoherent
dynamics)~\cite{weiss,lothar2}. Then, the exact dynamics approximately obeys
\begin{equation} \label{single-particle rate equation}
\frac{d\vec{P}(t)}{dt} = {\mathbf R}\, \vec{P}(t) \, ,
\end{equation}
where $\vec{P}$ collects the single particle populations and the matrix
${\mathbf R}$ contains the transition rates between adjacent sites. These rates
obey detailed balance reflecting the existence of a thermodynamic equilibrium
approached for long times, where the populations are Boltzmann distributed such
that $\mathbf{R}\, \vec{P}(t \rightarrow \infty) = 0$. However, as we have
shown in the previous sections, for the many-body time evolution this is no
longer true, and formulating Eq.~(\ref{single-particle rate equation}) with
$\vec{P}(t) = (P_{-S,-S}(t), P_{-S,-S+1}(t), \dots)$ and the corresponding
golden rule rates in $\mathbf{R}$ must inevitably fail to reproduce the correct
dynamics.

The idea is thus to start from the pseudo-angular momentum basis~(\ref{B_J})
and to employ separate rate descriptions for each invariant subspace with
initial and equilibrium populations according to Eqs.~(\ref{P_0})
and~(\ref{equil. pops B_J}), respectively, and with transition rates chosen
according to known expressions like golden rule
formulae~\cite{lothar1,lothar2}. The LMBS populations can then be obtained from
Eq.~(\ref{LMBS populations}). This approach, however, only applies for LMBS
populations for which the cross terms in Eq.~(\ref{LMBS populations}) vanish,
i.e.~if the initial density matrix $\rho(0)$ includes no finite off-diagonal
elements with respect to the basis $B_J$. In terms of the path-integral
picture, this restricts a rate approach to populations whose path-integral
expression~(\ref{decomposed path integral expression for A(t)}) collapses to a
single $A_{J_1,J_1}(t)$. For instance, for a system obeying Eq.~(\ref{J^2
 commutator}) and being prepared according to $P_{-S,-S}(0)=1$ or
$P_{S,S}(0)=1$, this holds for all LMBS populations, while for $P_{0,0}(0)=1$
only some of them can be reproduced this way
(cf.~Figs.~\ref{d_3_fully_degenerate} and~\ref{d_3_fully_degenerate_0,0}). Note
that Eq.~(\ref{J^2 commutator}) is no necessary condition for this approach, as
we could confirm for a system with $d=3$, $\epsilon_i=0$, $\Delta_1 =
\Delta_2$, $U=5\Delta_1$, and $P_{-1,-1}(0) = 1$, where despite a coupling
between $\mathcal{G}_0$ and $\mathcal{G}_1$ all LMBS populations with
$s^\downarrow + s^\uparrow \ne 0$ could be reproduced (not shown).

If, tough, off-diagonal terms are present in $\rho(0)$, they by no means can be
captured by a simple rate formalism, and the dynamics of the corresponding LMBS
populations will, at least for short to intermediate times, escape a
description along the lines of Eq.~(\ref{single-particle rate equation}). This
reflects the fact that, although unable to introduce transitions between them,
the phonon-induced coupling still correlates the dynamics on otherwise
disjunctive subspaces, an effect clearly beyond the scope of rate models. To
what extent the presence of non-diagonal elements of the density matrix do
influence the dynamics in specific situations, however, necessitates a deeper
analysis based on the exact path integral expression~(\ref{path integral
 expression for A(t)}) and a generalization of what is known about standard
spin-boson type of models~\cite{weiss}. Nevertheless, we heuristically found
that in the cases we investigated, LMBS populations which cannot be solely
expressed in terms of pseudo-angular momentum populations still can be well
approximated by
\begin{equation} \label{heuristic rate model}
\tilde{P}_{s^\downarrow, s^\uparrow}(t)
\equiv \bar{P}_{s^\downarrow, s^\uparrow}(t)
+ \left(1 - e^{-\gamma t}\right)
  \left(P^\infty_{s^\downarrow, s^\uparrow}
        - \bar{P}^\infty_{s^\downarrow, s^\uparrow}\right) \,,
\end{equation}
where $\bar{P}_{s^\downarrow, s^\uparrow}(t)$ denotes the corresponding LMBS
population as obtained from Eq.~(\ref{single-particle rate equation}), and
$\gamma$ is the smallest finite modulus of the eigenvalues of the rate matrices
of all invariant subspaces $\mathcal{G}_j$ which participate in the dynamics
(cf.~Fig.~\ref{d_3_fully_degenerate_0,0}). Whether Eq.~(\ref{heuristic rate
 model}) also covers a more general set of parameters, however, lays beyond
our knowledge as well as an explanation for observation that in cases where no
finite non-diagonal elements are present, Eq.~(\ref{heuristic rate model})
performs rather poorly.
Thus, Monte Carlo simulations are of particular importance to obtain insight
into the nonequilibrium dynamics of correlated charge transfer in cases where
symmetries are broken.

\section{Applications}
\label{Applications}

In this section we apply the general results of the previous sections to
specific models, particularly, to tight-binding systems with $d=3$ and $d=4$
sites. The simplest case $d=2$ is not explicitly addressed since it has been
extensively studied in Ref.~\cite{prl}. We only mention again that for this
generalization of the ordinary spin-boson model, a decomposition in
$\mathcal{G}^+ \equiv \mathcal{G}_1$ and $\mathcal{G}^- \equiv \mathcal{G}_0$
exists for \textit{arbitrary} system parameters, where the singlet state in the
latter subspace constitutes a DFS.

\subsection{The case $\mathbf{d=3}$}
\label{The case d 3}

\begin{figure}
\vspace*{0.7cm}
\epsfig{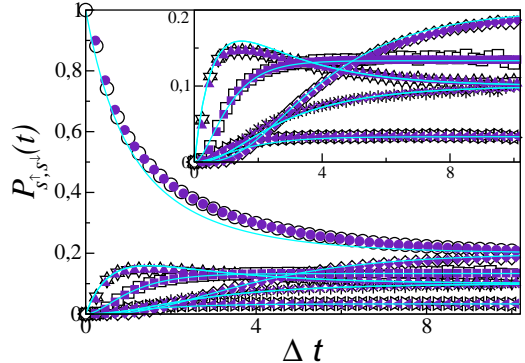}
\caption[]{\label{d_3_fully_degenerate} Initial dynamics of a two-particle
  system for $d=3$ with completely degenerate onsite energies and coupling
  strengths $\Delta_1 = \Delta_2 = \Delta$ and $U_C = 0$. Shown are PIMC data
  for the localized populations Eq.~(\ref{localized populations}) as obtained
  from Eq.~(\ref{path integral expression for A(t)}) (empty black symbols) and
  Eq.~(\ref{decomposed path integral expression for A(t)}) (filled gray
  symbols); gray lines denote results from the rate approach described in
  Sec.~\ref{Rate model}. Shown are $P_{-1,-1}$ (circles), $P_{-1,0}$
  (triangles up), $P_{-1,1}$ (triangles left), $P_{0,-1}$ (triangles down),
  $P_{0,0}$ (squares), $P_{0,1,}$ (``+''), $P_{1,-1}$ (triangles right),
  $P_{1,0}$ (``$\times$''), and $P_{1,1}$ (diamonds), with equilibrium values
  of 1/5, 1/10, 1/30, 1/10, 2/15, 1/10, 1/30, 1/10, and 1/5, respectively. The
  inset exhibits are more detailed view of the populations with no initial
  population. 
}
\vspace*{0.0cm}
\end{figure}

\begin{figure}
\vspace*{0.7cm}
\epsfig{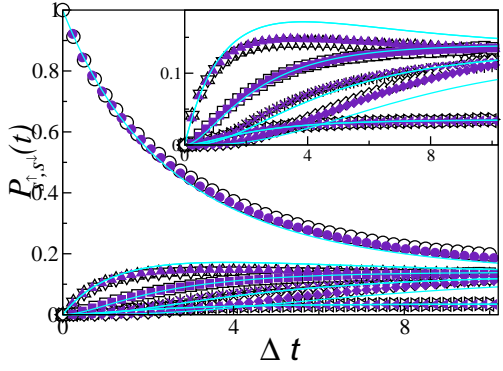}
\caption[]{\label{d_3__0_-2.5_0__5} Same as Fig.~\ref{d_3_fully_degenerate},
  but with $\epsilon_1 = \epsilon_3 = 0$, $\epsilon_2 = -2.5\Delta$, and $U_C =
  5\Delta$, with equilibrium values of 0.1269, 0.1343, 0.0349, 0.1343, 0.1394,
  0.1343, 0.0349, 0.1343, and 0.1269. }
\vspace*{0.0cm}
\end{figure}

As already stated above, for isotropic couplings and all on-site energies and
the Coulomb energy set to zero, one has a full decomposition of the Hilbert
space. Corresponding data for the dynamics of the localized
populations~(\ref{localized populations}) are depicted in
Figs.~\ref{d_3_fully_degenerate} and~\ref{d_3_fully_degenerate_0,0} for an
ohmic spectral density with exponential cutoff,
\begin{equation} \label{spectral density}
J(\omega)= 2 \pi \alpha \omega {\rm e}^{-\omega/\omega_c} \,,
\end{equation}
where $\alpha = 0.25$, $\omega_c = 5\Delta$, and $\hbar\beta = 0.1\hbar\Delta$
(the bath setup is the same for all populations shown here). Note the change in
the equilibrium populations due to the different initial preparations.
%
%
%
\begin{figure}
\vspace*{0.7cm}
\epsfig{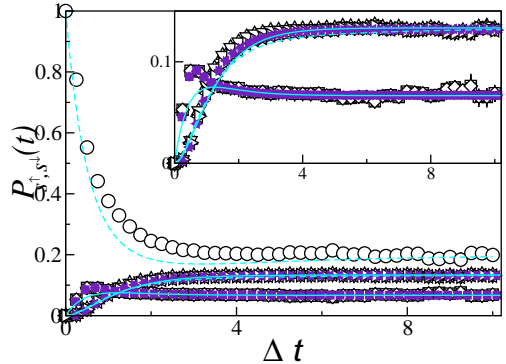}
\caption[]{\label{d_3_fully_degenerate_0,0} Same as
  Fig.~\ref{d_3_fully_degenerate}, but with initial preparation on $|0,0\rangle$. 
  Shown are $P_{-1,-1}$ (triangles left),
  $P_{-1,0}$ (squares), $P_{-1,1}$ (triangles up), $P_{0,-1}$ (``+''),
  $P_{0,0}$ (circles), $P_{0,1,}$ (``$\times$''), $P_{1,-1}$ (triangles down),
  $P_{1,0}$ (diamonds), and $P_{1,1}$ (triangles right), with equilibrium
  values of 2/15, 1/15, 2/15, 1/15, 1/5, 1/15, 2/15, 1/15, and 2/15,
  respectively. Dashed gray lines denote $\tilde{P}_{-1,1}(t)$,
  $\tilde{P}_{0,0}(t)$, and $\tilde{P}_{1,-1}(t)$~(\ref{heuristic rate
  model}).} 
\vspace*{0.0cm}
\end{figure}
%
Turning on the Coulomb interaction, i.e.~$U\neq 0$, a coupling between
$|\psi^{(0)}_0\rangle$ and $|\psi^{(2)}_0\rangle$ is introduced
(cf.~Fig.~\ref{d_3_invariant_subspaces}), with
$\langle\psi^{(2)}_0|H_S|\psi^{(0)}_0\rangle = (\sqrt{2}/3) U$, such that only
two invariant subspaces remain, i.e.
\begin{equation} 
\mathcal{H}_{d=3} = \mathcal{G}_{0\cup2}\cup \mathcal{G}_1
\,,
\end{equation}
with $\mathcal{G}_{0\cup2} = \mathcal{G}_0 \cup \mathcal{G}_2$. However, when
fixing the interaction strength to the specific value $U_C = \epsilon_{-1} +
\epsilon_1 - 2 \epsilon_0$, a full decomposition is recovered, see
Fig.~\ref{d_3__0_-2.5_0__5}.
%
%

\begin{figure}
\vspace*{0.7cm}
\epsfig{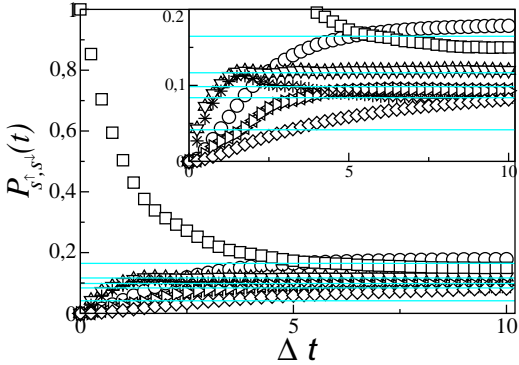}
\caption[]{\label{d_3__broken_symmetry} Same as
  Fig.~\ref{d_3_fully_degenerate}, but with $\Delta_2 = 0.7\Delta_1$,
  $\epsilon_1 = -1.65\Delta_1$, $\epsilon_2 = 0$, $\epsilon_3 = 3.5\Delta_1$,
  and $U_C = 5.1$, with equilibrium values of 0.1884, 0.1330, 0.0812, 0.1330,
  0.1567, 0.0795, 0.0812, 0.0795, and 0.0673. Gray lines here denote
  equilibrium populations as obtained from a Boltzmann distribution if all
  symmetries were broken.} 
\vspace*{0.0cm}
\end{figure}

Note that, without an external magnetic field, even for arbitrary system
parameters the particle-exchange symmetry cannot be broken: Albeit
$|\psi^{(0)}_0\rangle$ then typically couples to $|\psi^{(2)}_-1\rangle$,
$|\psi^{(2)}_0\rangle$, and $|\psi^{(2)}_1\rangle$, no couplings between
angular-momentum states with even and odd $J$ exist, such that \textit{always}
a non-Boltzmann distribution for the equilibrium populations is obtained
(cf.~Fig.~\ref{d_3__broken_symmetry}).
%
%

\subsection{The case $\mathbf{d=4}$: Example for a partial decomposition}
\label{The case d=4: Example for a partial decomposition}

\begin{figure}
\vspace*{0.7cm}
\epsfig{file=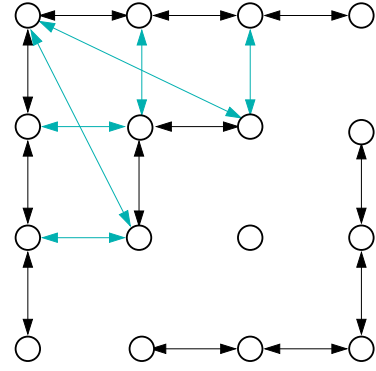, height=4.8cm}
\caption[]{\label{3} 2d mapping for an energetically degenerate $d=4$ chain. 
  Black solid lines: Couplings for $\Delta_3 = \Delta_1$, $\Delta_2 =
  2/\sqrt{3}\; \Delta_1$; gray lines: Additional couplings for $\Delta_3 =
  \Delta_2 = \Delta_1$.} 
\vspace*{0.0cm}
\end{figure}

For a tight-binding lattice with $d=4$ sites, a full decomposition of the
Hilbert space is already not possible for vanishing energies and isotropic
hopping terms. Nevertheless, one can still find a \textit{partial
 decomposition}
\begin{equation} 
\mathcal{H}_{d=4} = \mathcal{G}_0 \cup \mathcal{G}_2 \cap
\mathcal{G}_{1\cup3} \,,
\end{equation}
where $\mathcal{G}_{1\cup3} = \mathcal{G}_1 \cup \mathcal{G}_3$, and
$\mathcal{G}_0$, $\mathcal{G}_2$, and $\mathcal{G}_{1\cup3}$ are again
disjunctive and invariant with respect to $H$ (see Fig.~\ref{3}).
A Boltzmann distribution again can be observed for the equilibrium distribution
of $|\psi^{(j)}_m\rangle$ states for each invariant subspace but not for the
total Hilbert space. Contrary to $\mathcal{G}_1$ and $\mathcal{G}_3$, a
combination of invariant subspaces like $\mathcal{G}_{1\cup3}$ will in general
not exhibit a linear structure with respect to the tunnel couplings. For a
energetically non-degenerate systems, further couplings between $\mathcal{G}_0$
and $\mathcal{G}_2$ are introduced. However, even for arbitrary on-site
energies $E_i$, Coulomb energies $U$, and tunnel couplings $\Delta_i$ the
decomposition of the Hilbert space can never be completely destroyed due to the
underlying spin-permutation symmetry discussed in Sec.~\ref{Particle-exchange
 symmetry}.
%
%

\subsection{Quasi-invariant subspaces}
\label{Quasi-invariant subspaces}

For systems with a high symmetry (e.g.~when Eq.~(\ref{J^2 commutator})
applies), even a minute change in the system's parameters can lead to a
symmetry breaking. Along goes a modification of the structure of the Hilbert
space: Formerly invariant subspaces merge such that their number decreases.
Since the size of the new invariant subspaces exceeds that of the former ones,
the equilibration process within such a larger subspace is expected to slow
down accordingly. If, however, the coupling between formerly invariant
subspaces remains rather small, the relaxation process may in fact occur on two
(or more) time scales: One (or more) governing the internal dynamics within
each former subspace and another one, strongly separated from those, governing
the overall approach to equilibrium. In this situation, one may still speak of
\textit{quasi-invariant} subspaces which establish quasi-stationary states on
an intermediate time scale.

\begin{figure}
\vspace*{0.7cm}
\epsfig{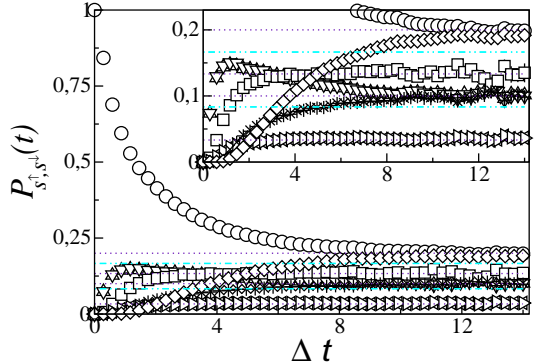}
\caption[]{\label{d_3_quasi-equilibrium} Same as in
  Fig.~\ref{d_3_fully_degenerate}, but with $U_C = 0.05\Delta$. The gray dot-dashed
  lines show the equilibrium populations $P^\infty_{-1,-1} = P^\infty_{1,1} =
  0.1663$ (circles and diamonds), $P^\infty_{-1,0} = P^\infty_{0,-1} =
  P^\infty_{0,1} = P^\infty_{1,0} = 0.0835$ (triangles up, triangles down,
  ``+'', and ``$\times$''), 0.0834 (triangles left and triangles right), and
  0.1666 (squares), respectively. Solid dark-gray dotted lines represent the
  equilibrium populations for a corresponding system with $U_C = 0$. }
\vspace*{0.0cm}
\end{figure}

For example, a weak particle interaction with $U_C = 0.05\Delta$ in an
otherwise energetically degenerate system with $d = 3$ leads to a weak coupling
between $|\psi^{(0)}_0\rangle$ and $|\psi^{(2)}_0\rangle$
(cf.~Fig.~\ref{d_3_invariant_subspaces}). Consequently, instead of the formerly
invariant subspaces $\mathcal{G}_0$, $\mathcal{G}_1$, and $\mathcal{G}_2$, now
only $\mathcal{G}_1$ and $\mathcal{G}_{0 \cup 2} = \mathcal{G}_0 \cup
\mathcal{G}_2$ remain invariant with respect to $H$. The smallest eigenvalue of
the according rate matrix $\mathbf{R}$ (see Sec.~\ref{Rate model}), describing
the long-time dynamics~\cite{lothar2}, then becomes $0.010\Delta$, suggesting
an equilibration timescale of the order of $100\Delta^{-1}$. The second
smallest eigenvalue, however, exceeds the latter by a factor of more than 20.
Accordingly, for $1 \ll \Delta t \ll 100$, the system will be trapped in a
'quasi-equilibrium', which still reflects the full symmetry and therefore
significantly differs from the true equilibrium state
(cf.~Fig.~\ref{d_3_quasi-equilibrium}). From an experimental point of view,
this quasi-equilibrium might easily be mistaken as the true equilibrium since
on the one hand, the equilibration timescale can take very large values for
systems with only slightly broken symmetries, while on the other such a
symmetry breaking is rather hard to identify since it can easily occur at
rather `harmless' system parameters. Note that this phenomenon can also occur
when applying a weak magnetic field which breaks the spin-permutation symmetry.

%
\subsection{Particle interaction}
\label{Particle interaction}

\begin{figure}
\vspace*{0.7cm}
\epsfig{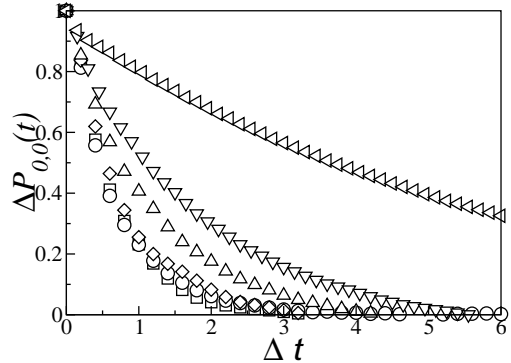}
\caption[]{\label{Coulomb_interaction} $\Delta P_{0,0}(t)$ for a system with
  $d=3$, $\epsilon_i = 0$, and $U_C/\Delta = 0$ (circles), 2.5 (squares), 5
  (diamonds), 7.5 (triangles up), 10 (triangles down), 15 (triangles left), and
  20 (triangles right), initially
  prepared in $|0,0\rangle$.} 
\end{figure}

An unique feature of many-particle systems is the interaction between its
constituents, where the Coulomb interaction is the most prominent example. In
our model~(\ref{H_S})--(\ref{H_I}), Coulomb interaction between particles
manifests itself as change of the energies of the localized states and
influences the dissipative system in quite different ways: On the one hand, it
can facilitate or break the symmetries discussed in
Secs.~\ref{Particle-exchange symmetry} and~\ref{Pseudo-angular momentum basis
 and invariant subspaces} and accordingly the decomposition of the full
Hilbert space into invariant subspaces, as has been intensively discussed
above. On the other, as has been already pointed out by
R.~Marcus~\cite{marcus}, a different energy offset between adjacent sites
$\Delta E_{j} = E_j-E_{j+1}$ introduces a phonon \textit{activation barrier}
between these sites, which is a unique signature of the interaction between
system and phonon bath. It reflects the fact that for a transition between
adjacent sites energy fluctuations are necessary for the reorganization of the
bath degrees of freedom after the tunneling of an electron. This latter process
can exhibit a strong impact, since for larger Coulomb interaction $U_j$, the
phonon activation barrier exceeds the electronic offset $\Delta E_j$ by far. In
fact, it can lead to rather counter-intuitive dynamics: If, for example, two
electrons are initially prepared on the same site of the chain, any Coulomb
interaction is expected to speed up the initial state's depopulation. However,
in competition to that the phonon interaction also generates an activation
barrier, which in turn slows down any transitions between the initial and
neighboring states. Therefore, only sufficiently weak Coulomb repulsions will
indeed lead to a faster emptying of the initial state; large repulsion
strengths, in contrast, cause a slow-down of the initial depopulation dynamics
and can even lead to a stabilization of the initial state.
Figure~\ref{Coulomb_interaction} visualizes this by comparing the initial
dynamics of systems with varying interaction strengths $U_C$. Note that the
turnaround takes places at around $U_C = 2.5\hbar\Delta$, which coincides with
the classical reorganization energy $\hbar\Lambda^{(cl)} = (1/\pi)
\int_0^\infty\!d\omega\, J(\omega)/\omega$.
%
%

We note in passing that a quantitative discussion of the effect of
phonon-induced interactions via the common environment is, unlike to the case
of two indistinguishable particles, not easy: While in the latter case one
could simply compare the dynamics of two particles on a 1d chain with Coulomb
repulsion $U_C$ with the QMC data for the dynamics of a single particle on a
corresponding 2d lattice~\cite{chemphys}, for two distinguishable particles
this approach is of no avail due to the intrinsically completely different
structures of the Hilbert spaces.

\section{Conclusions}
\label{Conclusions}

In this paper we have analysed the correlated nonequilibrium dynamics of two
interacting distinguishable fermions on a one-dimensional tight binding lattice
embedded into a phononic environment. As a major result, we have revealed the
crucial role of symmetries and provided their full characterization. In
particular, there is: (i) a spin-permutation symmetry which is fundamental in
that it is conserved independent of system and bath parameters; it leads to a
separation of the full Hilbert space into two disjunct subspaces. (ii) A
conservation of a total quasi-angular momentum $\vec{J}^2$ corresponding to
$[H,\vec{J}^2]=0$ for certain topologies of system parameters. This symmetry
determines a set of $J=d$ ($d$ number of sites) disjunct subspaces. The
dissipative many-body dynamics happens to occur in each of these subspaces
independently with transitions between them only mediated by the initial
preparation and/or the measurement observable. Recently, similar kinds of
symmetries have been discussed for quantum state transfer along networks as
realized e.g.~in linear arrays of Josephson junctions~\cite{bose}. In any case,
the time evolution in presence of the heat bath is not ergodic moving towards a
stationary state which is not Boltzmann like distributed. Of particular
relevance is the DFS, a subspace with no coupling to the bath at all, the
initial state of which is preserved for all times. For $d=2$ the nature of the
many body DFS is equivalent to the DFS discussed in the context of quantum
information processing for interacting spin-$1/2$ systems, but differs from the
latter for all $d>2$.

A direct practical consequence of the symmetry properties is that the dynamical
sign problem notoriously occurring in real-time quantum Monte Carlo approaches
can be substantially damped by performing the sampling in each of the subspaces
separately. Further, a rate description developed for single particle tight
binding systems and known to be applicable whenever the dynamics is
sufficiently incoherent, could be generalized to capture also the many body
relaxation. This approximation is of great importance as it not only allows to
obtain a deeper physical insight into the numerical QMC data, but also to
investigate the long-time domain which is typically not accessible by the PIMC.

We have illustrated these findings for one-dimensional Hubbard chains with
sites of up to $d=4$ and considerably long times, which may be also of
experimental interest. In particular, small arrays of quantum dots would be
appropriate candidates, where the above symmetries could directly be exploited
to control dynamical features of two electron transport. In fact, a fully
controllable device consisting of two coupled quantum dots has already been
realized~\cite{doubledot}. An additional external magnetic field further opens
a way to break or restore the fundamental spin-permuation symmetry. Note that
although here we have been mostly concerned with incoherent dynamics, the above
findings also apply to the domain of coherent transport as shown for $d=3$ in
Ref.~\cite{prl}.

Systems with more than two electrons or systems with indistinguishable
particles can be analysed along the same lines as above, however, with
differences in detail. In this context the case of many boson systems coupled
to a dissipative bath, i.e.~a bosonic Hubbard model with dissipation, is
certainly of great interest, e.g.~for condensed ultra-cold atomic gases.
Recently, we have already shown that for a two boson system a DFS exists only
for an odd number of lattice sites~\cite{chemphys}. Work to extend this study
to a full characterization of the symmetry properties as well as a
generalization of the rate description are currently in progress.

\acknowledgements We thank A. Komnik and H. Grabert for fruitful discussions.
This work has been supported by grants from the Ministry of Science, Research
and Arts of Baden W\"urttemberg No.~24-7532.23-11-11/1 and from the DFG. JA
acknowledges a Heisenberg fellowship of the DFG.

\appendix
\section{Discretized path integral for many-body systems}
\label{Appendix}

The derivation of a discretized version of the path-integral
expression~(\ref{path integral expression for A(t)}) is very similar to the
corresponding case of a single particle (see, e.g.,
Refs~\cite{weiss,egger,lothar1,lothar2}). Therefore, we here restrict ourselves
mainly to details specific to the many-particle case.

We begin with inserting $2q$ complete sets of projectors, one after each
short-time step $\tau \equiv t/q$,
\begin{eqnarray} \label{discretization}
A(t) &=& {\rm tr}\{W(t) A\}
\nonumber\\
&=& \sum_{|s^\downarrow_j|,|s^\uparrow_j| \le S}
\left(\prod_{j=2}^{2q} \int\!d\vec{x}_j\right) \langle
\vec{s}_1,\vec{x}_1|W(0)|\vec{s}_2,\vec{x}_2\rangle
\nonumber\\
&&\times \prod_{j=3}^q
 \langle \vec{s}_{j-1},\vec{x}_{j-1}|e^{i\tau H/\hbar}|\vec{s}_j,\vec{x}_j\rangle
\nonumber\\
&& \times \langle
\vec{s}_q,\vec{x}_q|A|\vec{s}_{q+1},\vec{x}_{q+1}\rangle
\nonumber\\
&& \times\prod_{j=q+2}^{2q} \langle
\vec{s}_{j-1},\vec{x}_{j-1}|e^{-i\tau
H/\hbar}|\vec{s}_j,\vec{x}_j\rangle \,,
\end{eqnarray}
where $|\vec{x}\rangle$ is a suitable basis of the environment, and $\vec{s}
\equiv (s^\downarrow, s^\uparrow)$ denotes the system's coordinates with
respect to $\mathcal{B}_{\rm loc}$~(\ref{B_loc}). Exploiting the symmetric
Suzuki splitting,
\begin{eqnarray} \label{Suzuki}
\lefteqn{\langle \vec{s}_{j-1},\vec{x}_{j-1}|e^{i\tau
    H/\hbar}|\vec{s}_j,\vec{x}_j\rangle} \nonumber\\
&=& \langle \vec{s}_{j-1},\vec{x}_{j-1} | e^{i\tau(H_I +
H_B)/2\hbar}
\nonumber\\
&&\times  e^{i\tau H_S/\hbar} e^{i\tau(H_I + H_B)/2\hbar}
| \vec{s}_j,\vec{x}_j
\rangle + O(\tau^3) \nonumber\\
&=& \exp\!\left\{ \frac{ia^2\tau}{\hbar}
        \left[(\vec{s}_{j-1}\cdot\vec{e})^2 + (\vec{s}_{j-1}\cdot\vec{e})^2\right]
              \sum_\alpha \frac{c_\alpha^2}{2m_\alpha \omega_\alpha^2}
          \right\}
\nonumber\\
&& \times K(\vec{s}_{j-1},\vec{s}_j)
\nonumber\\
&& \times \langle\vec{x}_{j-1} |
 \exp\!\left\{ \frac{i\tau}{2\hbar}
     \left[ -a(\vec{s}_{j-1} \cdot \vec{e}) \sum_\alpha c_\alpha X_\alpha
            + H_B\right]\right\}
\nonumber\\
&& \quad \times \exp\!\left\{ \frac{i\tau}{2\hbar}
     \left[ -a(\vec{s}_j \cdot \vec{e}) \sum_\alpha c_\alpha X_\alpha
            + H_B\right]\right\}|\vec{x}_j\rangle
\nonumber\\
&&{} + O(\tau^3) \,,
\end{eqnarray}
where
\begin{equation} 
K(\vec{s}_{j-1},\vec{s}_j) \equiv
 \langle\vec{s}_{j-1}|e^{i\tau H_S/\hbar}|\vec{s}_j\rangle
\end{equation}
denotes the free system's propagator over a time interval $\tau$, one obtains
an expression where the environmental degrees of freedom can be integrated out
exactly due to the harmonic nature of $H_B$ and $H_I$, leading to a discretized
version $\Phi[\{\vec{s}_j\}]$ of the influence functional~(\ref{influence
exponent}). Since, after replacing $\vec{s}_j \cdot \vec{e}$ with the
corresponding single-particle coordinate, the environmental terms in
Eq.~(\ref{Suzuki}) are then identical to those of the single-particle case
(cf., e.g., Ref~\cite{egger,lothar1,lothar2}), $\Phi[\{\vec{s}_j\}]$ can be
easily obtained from the single-particle expression without the need of redoing
the tedious integrals. Rewriting the system's coordinates with the aid of the
forward and backward paths introduced in Sec.~\ref{Path integral
representation},
\begin{eqnarray} 
s^{\downarrow/\uparrow}_j &\rightarrow&
s^{\downarrow/\uparrow}_j         \quad {\rm for}~1   \le
j \le q+1 \,, \\\nonumber s^{\downarrow/\uparrow}_j
&\rightarrow& \tilde{s}^{\downarrow/\uparrow}_{2q+2-j}
\quad {\rm for}~q+2 \le j \le 2q+1 \,,
\end{eqnarray}
the exponent of the influence functional becomes, in terms of the sum and
difference coordinates $\vec{\eta}_j \equiv (s^\downarrow_j +
\tilde{s}^\downarrow_j,s^\uparrow_j + \tilde{s}^\uparrow_j)$ and $\vec{\xi}_j
\equiv (s^\downarrow_j - \tilde{s}^\downarrow_j,s^\uparrow_j -
\tilde{s}^\uparrow_j)$, respectively,
\begin{eqnarray} \label{discretized influence exponent}
\Phi[\{\vec{\eta}_j, \vec{\xi}_j\}] &=& i \Big(\eta(0)
\cdot \vec{e}\Big)
     \sum_{j = 2}^q \left(\vec{\xi}_j \cdot \vec{e}\right) \hat{X}_j
\nonumber\\
&&{} + \sum_{j \ge k = 2}^q \left\{\left(\vec{\xi}_j \cdot
\vec{e}\right) \Lambda_{j-k}
\left(\vec{\xi}_k \cdot \vec{e}\right)\right. \nonumber\\
&&\left.{} + i \left(\vec{\xi}_j \cdot \vec{e}\right)
X_{j-k} \Big(\vec{\eta}_k \cdot \vec{e}\Big)\right\} \,. 
\end{eqnarray}
The real valued \dots $\Lambda_j$, $X_j$, and $\hat{X}_j$ are calculated
according to
\begin{eqnarray} 
\Lambda_0 + i X_0 &=& Q(\tau) \,,\nonumber\\
\Lambda_l + i X_l &=& Q((l-1)\tau) + Q((l+1)\tau) - 2Q(l\tau) \nonumber\\
&& {\rm for}\ 1 \le l \le q-2 \,,\nonumber\\
\hat{X}_j &=& {\rm Im}\{Q((j-2)\tau) - Q((j-1)\tau)\}\nonumber\\
&& {\rm for}\ 2 \le j \le q \,,
\end{eqnarray}
where
\begin{eqnarray}
Q(t) &=& \frac{1}{\pi} \int_0^\infty\!d\omega
\frac{J(\omega)}{\omega^2}
\{\coth(\hbar\beta\omega/2)[1-\cos(\omega t)] \nonumber\\
&&\qquad\qquad {} + i\sin(\omega t)\}\label{L(t)}
\end{eqnarray}
is the twice integrated bath-autocorrelation function $L(t)$~\cite{weiss}. We
thus arrive, as a discretized version of Eq.~(\ref{path integral expression for
A(t)}), at
\begin{eqnarray} \label{discretized path integral expression for A(t)}
\langle A(t)\rangle &=& \sum_{\{\vec{\eta}_j,
\vec{\xi}_j\}} a(\vec{\eta}_{q+1}) \prod_{j=2}^q
K(\vec{\eta}_j, \vec{\xi}_j; \vec{\eta}_{j+1},
\vec{\xi}_{j+1})
 \nonumber\\
&& \qquad \times \exp\!\left\{-\Phi[\{\vec{\eta}_j,
\vec{\xi}_j\}]\right\} \,,
\end{eqnarray}
which can now readily evaluated with PIMC schemes. 

\bibliography{general_Hubbard_chain}

\end{document}